\documentclass[preprintnumbers,article,amsmath,amssymb,floatfix,10pt,prd,onecolumn,
superscriptaddress,nofootinbib]{revtex4-2}
\usepackage{titlesec}

\titleformat*{\section}{\LARGE\bfseries}
\titleformat*{\subsection}{\Large\bfseries}
\titleformat*{\subsubsection}{\large\bfseries}
\titleformat*{\paragraph}{\large\bfseries}
\titleformat*{\subparagraph}{\large\bfseries}
\usepackage{bm}
\usepackage{amsfonts}
\usepackage{latexsym}
\usepackage[latin1]{inputenc}
\usepackage{graphicx}
\usepackage{amsmath}
\usepackage{mathtools} 
\usepackage{palatino}
\usepackage{mathpazo}
\usepackage{textcomp}
\usepackage{float}
\usepackage{booktabs}
\usepackage{dcolumn}
\usepackage{ragged2e}
\usepackage{hyperref}
\hypersetup{colorlinks,citecolor=blue}
\hypersetup{colorlinks=true,linkcolor=brown,filecolor=magenta,    urlcolor=blue}
\usepackage{amsthm}
\usepackage{xcolor}
\usepackage{orcidlink}
\usepackage{epsfig}
\usepackage[caption=false]{subfig}
\usepackage{commath}
\usepackage{cancel, ulem}

\usepackage{multirow}
\newcommand{\m}{\mathring}
\newcommand{\be}{\begin{equation}}
\newcommand{\ee}{\end{equation}}
\newcommand{\bea}{\begin{eqnarray}}
\newcommand{\eea}{\end{eqnarray}}
\newcommand{\eeas}{\end{eqnarray*}}
\newcommand{\beas}{\begin{eqnarray*}}

\def\jnl@style{\it}
\def\aaref@jnl#1{{\jnl@style#1}}

\def\aaref@jnl#1{{\jnl@style#1}}

\def\aj{\aaref@jnl{AJ}}                   
\def\apj{\aaref@jnl{ApJ}}                 
\def\apjl{\aaref@jnl{ApJ}}                
\def\apjs{\aaref@jnl{ApJS}}               
\def\apss{\aaref@jnl{Ap\&SS}}             
\def\aap{\aaref@jnl{A\&A}}                
\def\aapr{\aaref@jnl{A\&A~Rev.}}          
\def\aaps{\aaref@jnl{A\&AS}}              
\def\mnras{\aaref@jnl{Mon.~Not.~Roy.~Astron.~Soc.}}             
\def\prd{\aaref@jnl{Phys.~Rev.~D}}        
\def\prc{\aaref@jnl{Phys.~Rev.~C}}  
\def\prl{\aaref@jnl{Phys.~Rev.~Lett.}}    
\def\qjras{\aaref@jnl{QJRAS}}             
\def\skytel{\aaref@jnl{S\&T}}             
\def\ssr{\aaref@jnl{Space~Sci.~Rev.}}     
\def\zap{\aaref@jnl{ZAp}}                 
\def\nat{\aaref@jnl{Nature}}              
\def\aplett{\aaref@jnl{Astrophys.~Lett.}} 
\def\apspr{\aaref@jnl{Astrophys.~Space~Phys.~Res.}} 
\def\physrep{\aaref@jnl{Phys.~Rep.}}      
\def\physscr{\aaref@jnl{Phys.~Scr}}       
\def\commat{\aaref@jnl{Comm.~Math.~Phys.}}              
\def\science{\aaref@jnl{Science}}               
\def\cqg{\aaref@jnl{Classical Quant.~Grav.}}            
\def\jpcs{\aaref@jnl{JPCS}}                                     
\def\ijmpd{\aaref@jnl{Int.~J.~Mod.~Phys.~D}}                    
\def\grg{\aaref@jnl{Gen.~Relat.~Gravit.}}               
\def\rpp{\aaref@jnl{Rep.~Prog.~Phys.}}          
\def\npa{\aaref@jnl{Nucl.~Phys.~A}}        
\def\lrr{\aaref@jnl{Living Rev.~Rel.}}                   
\def\jcap{\aaref@jnl{J.~Cosmology Astropart.~Phys.}}    
\def\rmp{\aaref@jnl{Rev.~Mod.~Phys.}}   
\def\epjc{\aaref@jnl{Eur.~Phys.~J.~C}} 
\def\plb{\aaref@jnl{~Phy.~Lett.~B}} 
\def\mpla{\aaref@jnl{Mod.~Phy.~Lett.~A}} 
\def\arxiv{\aaref@jnl{arxiv.org}}

\allowdisplaybreaks[1]

\begin{document}

\title{Energy condition bounds on $f(Q)$ model parameters in a curved FLRW Universe }

\author{Ganesh Subramaniam\orcidlink{}}
\email{ganesh03@1utar.my}
\address{Department of Mathematical and Actuarial Sciences\\
Universiti Tunku Abdul Rahman, Jalan Sungai Long, 43000 Cheras, Malaysia}

\author{Avik De\orcidlink{0000-0001-6475-3085}}
\email{avikde@utar.edu.my}
\address{Department of Mathematical and Actuarial Sciences\\
Universiti Tunku Abdul Rahman, Jalan Sungai Long, 43000 Cheras, Malaysia}

\author{Tee-How Loo\orcidlink{}}
\email{looth@um.edu.my}
\address{Institute of Mathematical Sciences, Faculty of Science\\
Universiti Malaya, 50603 Kuala Lumpur, Malaysia}

\author{Yong Kheng Goh\orcidlink{}}
\email{gohyk@utar.edu.my}
\address{Department of Mathematical and Actuarial Sciences\\
Universiti Tunku Abdul Rahman, Jalan Sungai Long, 43000 Cheras, Malaysia}

\footnotetext{The research was supported by the Ministry of Higher Education (MoHE), through the Fundamental Research Grant Scheme (FRGS/1/2021/STG06/UTAR/02/1). }

\begin{abstract}
In this exclusive study of the modified $f(Q)$ theory of gravity in the open and closed type Friedmann-Lema\^itre-Robertson-Walker (FLRW) Universe model, we impose some constraints from the classical energy conditions. The viable range of parameter $\beta$ for two different $f(Q)$ models, $f(Q)=Q+\beta Q^2$ and $f(Q)=Q+\beta\sqrt{-Q}$, are analyzed in details and the related cosmological implications are discussed. Violation of effective strong energy condition is resulting into late-time acceleration of the Universe. Present observational values of Hubble parameter and deceleration parameter are used to constrain the parameters.  
\end{abstract}
\maketitle
\section{Introduction}
By imposing non-negativity on local energy, singularity at the beginning of the Universe, and the causal structure of the Universe, Penrose and Hawking initially popularised energy condition (EC) in the framework of the classical general relativity in order to comprehend the singularity generated by gravitational collapse \cite{penrose1965, hawking1966, hawking1966i, hawking1966ii,hawking1969,hawking1970}. Ordinary matter do not satisfy observational data for the accelerated expansion of the Universe \cite{ries1998, perlmutter1999} which urges the introduction of alternative energy, the so-called dark energy (DE). All these model-independent assessments agree that DE with negative effective pressure dominates the present cosmic fluid. Unfortunately, all attempts to physically detect DE in the cosmos went in vain so far. This resulted in the birth of alternatives curvature based gravity theories. $f(R)$-gravity, $f(R,T)$-gravity, $f(R,G)$-gravity and etc were introduced to provide an explanation of these problems geometrically without the need for the dark sector \cite{felice2010}. Metric teleparallel equivalent of GR (TEGR) was introduced by Einstein himself \cite{1} and it is studied widely where the ``metric-compatible and torsion-free'' Levi Civita connection is replaced by torsion based teleparallel connection. An extension in the form of $f(\mathbb{T})$ theory in the metric teleparallelism was introduced to tackle the dark sector \cite{fT1st,fT}. There is another kind of teleparallel theory available in the literature, the symmetric teleparallelism \cite{Nester}. In the present study, we concentrate on the extended $f(Q)$ theories of gravity, newly-proposed \cite{coincident} in the symmetric teleparallelism to avoid the DE-dependencies. It is customary to analyze the ECs in any proposed theory of modified gravity. However, this is a delicate topic in the realm of beyond-GR scenarios, as discussed in \cite{ecfr/2018,ec,ecgen}. There are several approaches in the literature deriving energy conditions in modified gravity, we can see for instance, EC constraints in $f(R)$ theory \cite{11, 12,bergliaffa2006}, $f(G)$ theory \cite{13, 14}, $f(T)$ theory \cite{15}, $f(G,T)$ theory \cite{16}, $f(R,T,R_{\mu\nu}T_{\mu\nu})$ theory \cite{17}, $f(R,G)$ theory \cite{18}, $f(R,\Box R,T)$ theory \cite{19}, $f(R,T)$ theory \cite{20}, among others.
 
 In the last couple of years, Several important publications came up on the $f(Q)$ gravity theory and its cosmological implications, see \cite{cosmology,cosmography,barros,lu,lcdm,lin,de/comment,de/iso,de/acc, de/complete,gde,ad/viability,cosmology_Q,redshift,signature,lcdm1,siren,recon,recon1,anisotropy,agrawal2023,maurya2022,narawade2022,narawade2023,narawade2023a,de/phase,de/probe} and the references therein. The corresponding ECs were also discussed \cite{fQec,fQec1}. However, except \cite{fQec1}, all these studies were solely carried out in the spatially-flat Friedmann-Lema\^itre-Robertson-Walker (FLRW) model of the Universe and the line element was specifically taken in Cartesian coordinates. Under this setting, a vanishing affine connection was used to formulate the $f(Q)$ theory, known as the coincident gauge choice. The whole formulation was simplified in this particular gauge as the covariant derivative reduced to partial derivative. However, we paid the price by forcing the $f(Q)$ theory to be equivalent to the $f(T)$ theory, producing an identical Friedmann type equations of pressure and energy density \cite{issueftfq}. In our last work \cite{fQec1}, we have alleviated this issue by considering a non-vanishing affine connection, however, it was still studied in the background of spatially flat FLRW spacetime. 
 
 In most of the cosmological works, researchers presume that the observable Universe is exactly spatially flat, that is, $k=0$. However, $k$ should have been constrained every time whenever the latest observational dataset is available. Therefore, ideally, we should incorporate the spatial curvature $k$ into account. There are some recent works where the effect of the spatial curvature were studied extensively \cite{yang2022,pan,holo,valentino2020,vagnozzi2021,vagnozzi2021a,dhawan2021,glanville2022}. Naturally, it is worthwhile to study the $f(Q)$ theory in the open and closed type FLRW model with $k=\pm 1$. The main challenge had been the complexity in the mathematical formalism of symmetric teleparallelism in such a background spacetime, and until \cite{FLRW/connection} there was not much attempt made to demonstrate the clear formulation in the open and closed type FLRW model.  

 The present article is organized as follows:\\After the introduction, in Section \ref{sec2} we provide the basic mathematical formalism of $f(Q)$ theory, followed by the construction of $f(Q)$ from a non-vanishing affine connection in a spatially curved (both positively as well as negatively) FLRW Universe in Section \ref{sec3}. Such connection coefficients involve a so-far unconstrained function of time, $\gamma(t)$. In our present study we consider two most prominent ansatz, a constant $\gamma=\gamma_0$ and $\gamma\propto a(t)$, the scale factor, motivated by the observation that in the latter case in negatively curved spacetime, the energy is conserved in a model-independent manner. The Friedmann-like equations of energy and pressure for ordinary matter and for effective counterparts are also provided. EC expressions corresponding to $f(Q)$ theory are presented in the brief Section \ref{sec4}. Next in Section \ref{sec5}, we do a model specific analysis for two $f(Q)$ models, $f(Q)=Q+\beta Q^2$ and $f(Q)=Q+\beta\sqrt{-Q}$, being done in two separate subsections \ref{model1} and \ref{model2}, respectively. For each $f(Q)$ model, positively and negatively curved spacetimes are discussed separately, and in each of these cases two ansatz are analyzed. Contour plots of all ECs are followed in each such ansatz, accompanied by two figures displaying the character of $p^{eff}$, and $\beta$ vs $\gamma$ for DEC, NEC and WEC, in each case. Finally, we conclude in Section \ref{sec6}.

\section{A briefing on symmetric teleparallelism}\label{sec2}
The symmetric teleparallel theory of gravity was formulated based on a general affine connection $\Gamma^\alpha_{\,\,\, \beta\gamma}$, defined by 
\begin{equation} \label{connc}
\Gamma^\lambda{}_{\mu\nu} = \mathring{\Gamma}^\lambda{}_{\mu\nu}+L^\lambda{}_{\mu\nu}
\end{equation}
with vanishing curvature and null torsion and in this theory we let the non-metricity of the underlying geometry controls the gravity. We first define the non-metricity tensor 
\begin{equation} \label{Q tensor}
Q_{\lambda\mu\nu} = \nabla_\lambda g_{\mu\nu} \,.
\end{equation}
The two possible traces of the non-metricity tensor are 
\[
Q_{\lambda}=Q_{\lambda\mu\nu}g^{\mu\nu}; \quad \tilde Q_{\nu}=Q_{\lambda\mu\nu}g^{\lambda\mu}.
\]
The disformation tensor $L^\lambda{}_{\mu\nu}$ and the superpotential tensor $P^\lambda{}_{\mu\nu}$ are respectively given by
\begin{equation} \label{L}
L^\lambda{}_{\mu\nu} = \frac{1}{2} (Q^\lambda{}_{\mu\nu} - Q_\mu{}^\lambda{}_\nu - Q_\nu{}^\lambda{}_\mu) \,.
\end{equation}
\begin{equation} \label{P}
P^\lambda{}_{\mu\nu} = \frac{1}{4} \left( -2 L^\lambda{}_{\mu\nu} + Q^\lambda g_{\mu\nu} - \tilde{Q}^\lambda g_{\mu\nu} -\frac{1}{2} \delta^\lambda_\mu Q_{\nu} - \frac{1}{2} \delta^\lambda_\nu Q_{\mu} \right) \,.
\end{equation} 
We consider non-metricity scalar  
\begin{equation} \label{Q}
Q=Q_{\lambda\mu\nu}P^{\lambda\mu\nu}= \frac{1}{4}(-Q_{\lambda\mu\nu}Q^{\lambda\mu\nu} + 2Q_{\lambda\mu\nu}Q^{\mu\lambda\nu} +Q_\lambda Q^\lambda -2Q_\lambda \tilde{Q}^\lambda).
\end{equation}

However, being equivalent to GR, the symmetric teleparallelism inherit the same `dark' problem as in GR, and so a modified $f(Q)$-gravity has been introduced in the same way as a modified $f(R)$-theory was introduced to extend GR. By varying the action term 
\begin{equation*}
S = \frac1{2\kappa}\int f(Q) \sqrt{-g}\,d^4 x
+\int \mathcal{L}_M \sqrt{-g}\,d^4 x
\end{equation*}
with respect to the metric to obtain the field equation in fully-covariant form \cite{zhao}
\begin{equation} \label{FE}
f_Q \m{G}_{\mu\nu}+\frac{1}{2} g_{\mu\nu} (f_QQ-f) + 2f_{QQ} P^\lambda{}_{\mu\nu} \m{\nabla}_\lambda Q = \kappa T^{m}_{\mu\nu}.
\end{equation}
We can rewrite (\ref{FE}) with $\kappa=1$ in an GR equivalent form
\begin{align}\label{FEeq}
    \m{G}_{\mu\nu}= \frac1{f_Q}T^{eff}_{\mu\nu}=\frac 1{f_Q}T^{m}_{\mu\nu}+T^{DE}_{\mu\nu}
\end{align}
where $T^{DE}_{\mu\nu}=\frac{1}{f_Q}\left[\frac{1}{2}g_{\mu\nu}(f-Qf_Q)-2f_{QQ}\mathring{\nabla}_\lambda QP^\lambda_{\mu\nu}\right]$ denotes the additional terms produced from the geometrical modification of the gravity theory in the present instance. We can very well visualise this as the component which works as some kind of fictitious dark energy. 

\section{The homogeneous and isotropic model of the Universe}\label{sec3}
The spatially curved homogeneous and isotropic FLRW spacetime metric is given by
\begin{align}
ds^2 = -\mathrm{d} t^2 
+a\left(t\right)^{2}\left( \frac{dr^2}{1-kr^2} +r^2\mathrm{d}\theta^2+r^2\sin^2\theta\mathrm{d} \phi^2\right), \quad k=\pm1
\end{align}
In this background spacetime, the compatible connection was discussed in \cite{FLRW/connection}
\begin{align} 
\Gamma^t{}_{tt}=&-\frac{k+\dot\gamma}\gamma, 
	\quad 					\Gamma^t{}_{rr}=\frac{\gamma}{1-kr^2}, 
	\quad 					\Gamma^t{}_{\theta\theta}=\gamma r^2, 
	\quad						\Gamma^t{}_{\phi\phi}=\gamma r^2\sin^2\theta								\notag\\
\Gamma^r{}_{tr}=&-\frac{k}{\gamma}, 
	\quad  	\Gamma^r{}_{rr}=\frac{kr}{1-kr^2}, 
	\quad		\Gamma^r{}_{\theta\theta}=-(1-kr^2)r, 
	\quad		\Gamma^r{}_{\phi\phi}=-(1-kr^2)r\sin^2\theta,												\notag\\
\Gamma^\theta{}_{t\theta}=&-\frac{k}{\gamma}, 
	\quad		\Gamma^\theta{}_{r\theta}=\frac1r,
	\quad		\Gamma^\theta{}_{\phi\phi}=-\cos\theta\sin\theta,										\notag\\
\Gamma^\phi{}_{t\phi}=&-\frac k\gamma, 
	\quad 	\Gamma^\phi{}_{r\phi}=\frac1r, 
	\quad 	\Gamma^\phi{}_{\theta\phi}=\cot\theta.
\end{align}

The corresponding non-metricity scalar $Q$  can be calculated from (\ref{Q})  as
\begin{equation}\label{Qf}
    Q(t)=-3\left[2H^2+\left(\frac{3k}{\gamma}-\frac{\gamma}{a^2}\right)H-\frac{2k}{a^2}-k\frac{\dot{\gamma}}{\gamma^2}-\frac{\dot{\gamma}}{a^2}\right].
\end{equation}
From the field equation (\ref{FE}) we obtain the Friedmann like equations
\begin{align}\label{rho}
\rho^{m}=&\frac12f+\left(3H^2+3\frac k{a^2}-\frac12Q\right)f_Q+\frac32\dot Q\left(-\frac k\gamma-\frac\gamma{a^2}\right)f_{QQ}.
\end{align}
\begin{align}\label{p}
p^{m}=&-\frac12f+\left(-3H^2-2\dot H-\frac k{a^2}+\frac12Q\right)f_Q
        +\dot Q\left(-2H-\frac32\frac k\gamma+\frac12\frac\gamma{a^2}\right)f_{QQ}.
\end{align}

The effective pressure and energy density can also be derived
\begin{align}\label{rho_eff}
\rho^{eff}=&\rho^m+\frac{1}{2}(Qf_Q-f)+\frac{3}{2}\dot{Q}f_{QQ}\left(
        \frac{\gamma}{a^2}+\frac{k}{\gamma}\right) .\\
   p^{eff}=&p^m-\frac{1}{2}(Qf_Q-f)
   -\frac12\dot{Q}f_{QQ}\left(\frac{\gamma}{a^2}-\frac{3k}{\gamma}-4H \right) .
\label{p_eff}
\end{align}


\section{Energy conditions}\label{sec4}

We consider a perfect fluid type ordinary matter whose stress-energy tensor $T^{m}_{\mu\nu}$ given by
\begin{align}
    T^{m}_{\mu\nu}=(p^{m}+\rho^{m})u_\mu u_\nu+p^{m}g_{\mu\nu}
\end{align} 
where $p^{m}$ and $\rho^{m}$ denote the pressure and energy density. In this particular form of stress-energy tensor, the four classical ECs take the following simple forms, 
\begin{itemize}
    \item Null energy condition (NEC): $\rho^{m}+p^{m}\ge0$.
    \item Weak energy condition (WEC): $\rho^{m}\ge0$ and $\rho^{m}+p^{m}\ge0$.
    \item Dominant energy condition (DEC): $\rho^{m}\pm p^{m}\ge0$. 
    \item Strong energy condition (SEC): $\rho^{m}+3p^{m}\ge 0$ and $\rho^{m}+p^{m}\ge0$.
\end{itemize}
Consequently, the ECs in this context are only a set of restrictions on the possible linear combinations of pressure and energy density. Ordinary matter always satisfies the WEC, NEC and SEC due to its positive pressure and energy density. In addition, we obtain a valid DEC (in the form of $\rho<p$) if we assume that the speed of sound in ordinary matter is always smaller than the speed of light. And this is gravity theory independent, so even if we modify the left hand side of the field equations in GR and consider the currently discussed $f(Q)$ gravity theory (\ref{FE}), the right hand side is still constrained by the set of EC equations above. Naturally, when we express this field equations (\ref{FE}) in the equivalent form (\ref{FEeq}), the $T^m_{\mu\nu}$ are still bound by the ECs, but the $T^{eff}_{\mu\nu}$ or $T^{DE}_{\mu\nu}$ are not. Therefore, we can utilise the ECs to find the suitable ranges of the model parameters and then examine the ``non-ordinariness" of the effective pressure and energy density in those regions.

By using equation (\ref{rho}) and (\ref{p}), we have the following expressions:
\begin{align}\label{aa}
    \rho^m=&\frac12f+\left(3H^2+3\frac k{a^2}-\frac12Q\right)f_Q+\frac32\dot Q\left(-\frac k\gamma-\frac\gamma{a^2}\right)f_{QQ}\\
    \rho^m-p^m=&f+\left(6H^2+2\Dot{H}-Q+\frac{4k}{a^2}\right)f_Q+2\dot{Q}\left(H-\frac{\gamma}{a^2}\right)f_{QQ}\\
    \rho^m+p^m=&-2\left(\Dot{H}-\frac{k}{a^2}\right)f_Q-\dot{Q}\left(2H+\frac{3k}{\gamma}+\frac{\gamma}{a^2}\right)f_{QQ}\\
    \rho^m+3p^m=&-f+(Q-6H^2-6\Dot{H})f_Q-6\dot{Q}\left(H+\frac{k}{\gamma}\right)f_{QQ}.\label{bb}
\end{align}

\section{Model specific analysis}\label{sec5}

\subsection{$f(Q)= Q+\beta Q^2$}\label{model1}
In this subsection, we consider the simplest extension of GR, in terms of the quadratic form of $f(Q)$. This particular model was deeply investigated in \cite{cosmology_Q} to compare with $\Lambda$CDM. Later the model was also used in \cite{de/probe} to study scalar field inflation, and in \cite{fQec,fQec1} to analyze energy conditions in $f(Q)$ theory, among others.
By engaging the deceleration parameter $q(t)=-\frac{\Ddot{a}}{aH^2}=-\left[1+\frac{\dot{H}}{H^2}\right]$, the expressions (\ref{aa})-(\ref{bb}) in terms of the present values $H_0$ and $q_0$ can be written as
\begin{align}
\rho^m& = 3\left(H_0^2+\frac{k}{a_0^2}\right)+6\beta Q_0\left(H_0^2+\frac{k}{a_0^2}\right)-\frac{1}{2}\beta Q_0^2-3\beta\dot{Q}_0\left(\frac{k}{\gamma}+\frac{\gamma}{a_0^2}\right) \label{wecp}\\
    \rho^m-p^m& = 2H_0^2(2-q_0)+\frac{4k}{a_0^2}-\beta Q_0^2+4\beta Q_0\left(H_0^2(2-q_0)+\frac{2k}{a_0^2}\right)+4\beta \dot{Q}_0\left(H_0-\frac{\gamma}{a_0^2}\right) \label{decp}\\
    \rho^m+p^m& = 2H_0^2(1+q_0)+\frac{2k}{a_0^2}+4\beta Q_0\left(H_0^2(1+q_0)+\frac{k}{a_0^2}\right)-2\beta\dot{Q}_0\left(2H_0+\frac{\gamma}{a_0^2}+\frac{3k}{\gamma}\right)\label{necq}\\
     \rho^m+3p^m& = 6H_0^2q_0+\beta Q_0^2+12\beta Q_0 H_0^2q_0-12\beta \dot{Q}_0\left(H_0+\frac{k}{\gamma}\right) \label{secp}
\end{align}
where the present value of the non-metricity scalar $Q_0$ produces
\begin{align*}
    \dot{Q}_0& = -3\left[\frac{4 k}{a_0^2}H_0-4H_0^3(1+q_0)-\left(\frac{3 k}{\gamma}-\frac{\gamma}{a_0^2}\right)H_0^2(1+q_0)+\frac{2 \dot{\gamma}}{a_0^2}H_0+\frac{2 k \dot{\gamma}^2}{\gamma^3}+H_0\left(\frac{2 \gamma} {a_0^2}H_0-\frac{\dot{\gamma}}{a_0^2}-\frac{3 k \dot{\gamma}}{\gamma^2}\right)-\frac{\ddot{\gamma}}{a_0^2}-\frac{k \ddot{\gamma}}{\gamma^2}\right]\label{Qd}\\
 Q_0^2&=9\left[4H_0^4+4\left(\frac{3k}{\gamma}-\frac{\gamma}{a_0^2}\right)H_0^3+\left(\frac{9k^2}{\gamma^2}-\frac{14k}{a_0^2}+\frac{\gamma^2}{a_0^4}-\frac{8k\dot{\gamma}}{\gamma^2}-\frac{8\dot{\gamma}}{a_0^2}\right)H_0^2-4\left(\frac{3k^2\dot{\gamma}}{\gamma^3}+\frac{2k\dot{\gamma}}{a_0^2\gamma}-\frac{\gamma\dot{\gamma}}{a_0^4}\right)H_0\right.\\&\left.+\frac{4k^2}{a_0^4}+\frac{k^2\dot{\gamma}^2}{\gamma^4}+\frac{2k\dot{\gamma}^2}{a_0^2\gamma^2}+\frac{\dot{\gamma}^2}{a_0^4}-\frac{8k^2\dot{\gamma}}{a_0^2\gamma^2}-\frac{8k\dot{\gamma}}{a_0^4}\right]
\end{align*}

whereby the observational values of $H_0=67.9 km s^{-1} Mpc^{-1}$ 
, $q_0= -0.55$ 
, respectively, can be utilised; $a_0$ can be taken as 1 for the present time. The effective EOS parameter $\omega^{eff}$ shows a constant value of $-0.7$. In this regard, it is important to take note of the strong discrepancy between local and early-time estimates of the Hubble constant $H_0$, termed as the $H_0$ tension, which could be pointing towards new physics beyond the concordance $\Lambda$CDM model \cite{H0}.
In the following we attempt to analyse few possible cases of $\gamma(t)$; without loss of generality, we assume it to be a positive function of time.
 \subsubsection{$\gamma(t)=\gamma_0$, a constant}
We need to analyse two separate situations, $k=1$ and $k=-1$, the closed and open Universe models, respectively.
\paragraph{Closed Universe model, $k=1$:}
The expressions (\ref{wecp})--(\ref{secp}) in this case reduce to
\begin{align}
    \rho^m=&\frac{\beta  (80912.7 \gamma_0^4+6.20077\times10^6\gamma_0^3-1.14748\times10^9\gamma_0^2-3.88780\times 10^7 \gamma_0-242738.0)}{\gamma_0^2}+13834.2\\
    \rho^m-p^m=&\frac{\beta  (94052.4 \gamma_0^4+4.88598\times 10^6 \gamma_0^3-1.60660\times 10^9 \gamma_0^2-5.74730\times 10^7 \gamma_0-373443.0)}{\gamma_0^2}+23517.1\\
    \rho^m+p^m=&\frac{\beta  (67773.0 \gamma_0^4+7.51557\times 10^6 \gamma_0^3-6.88364\times 10^8 \gamma_0^2-2.02830\times 10^7 \gamma_0-112033.0)}{\gamma_0^2}+4151.37\\
    \rho^m+3p^m=&\frac{\beta  (41493.7 \gamma_0^4+1.01452\times 10^7 \gamma_0^3+2.29871\times 10^8 \gamma_0^2+1.69070\times 10^7 \gamma_0+149377.0)}{\gamma_0^2}-15214.4.\label{seck=1}
\end{align}

 The coefficients of the model parameter $\beta$ are polynomials of $\gamma_0$ in each EC expression above, divided by $\gamma^2_0$. The polynomials have positive zeroes (at $86.8, 107.3$ and $59.6$, respectively) excluding (\ref{seck=1}). We investigate the environments around these zeroes as depicted in the figures followed; we also offer reasonable analysis for each of them below.

\begin{figure}[H]
 \begin{minipage}[b]{0.4\textwidth}
   \includegraphics[width=\textwidth]{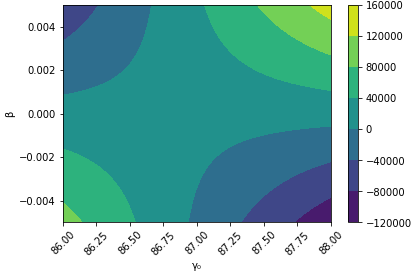}
   \caption{$\rho^m$ for $k =+1$}
 \end{minipage}
 \hfill
 \begin{minipage}[b]{0.4\textwidth}
   \includegraphics[width=\textwidth]{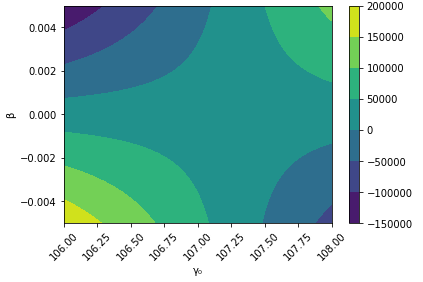}
   \caption{$\rho^m-p^m$ for $k =+1$}
 \end{minipage}
 
 \begin{minipage}[b]{0.4\textwidth}
   \includegraphics[width=\textwidth]{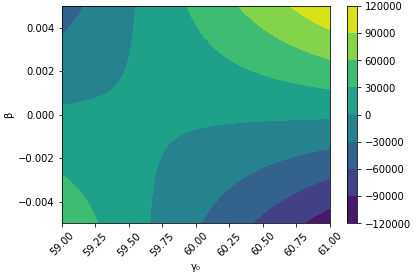}
   \caption{$\rho^m+p^m$ for $k=+1$}
 \end{minipage}
 \hfill
 \begin{minipage}[b]{0.4\textwidth}
   \includegraphics[width=\textwidth]{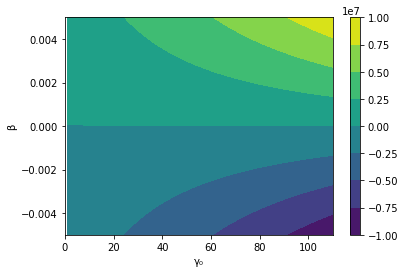}
   \caption{$\rho^m+3p^m$ for $k=+1$}
 \end{minipage}
 \end{figure}

 \begin{figure}[H]
  \begin{minipage}[b]{0.45\textwidth}
  \includegraphics[width=\textwidth]{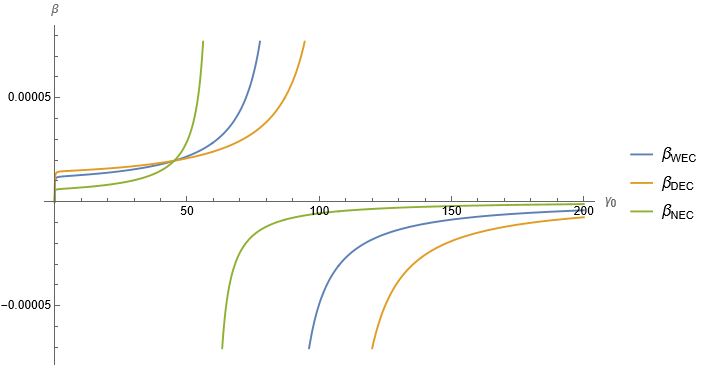}
   \caption{$\beta_{WEC},\beta_{DEC}, \beta_{NEC}$ vs $\gamma_0$ for $k =+1$}
 \end{minipage}
\hfill
 \begin{minipage}[b]{0.39\textwidth}
  \includegraphics[width=\textwidth]{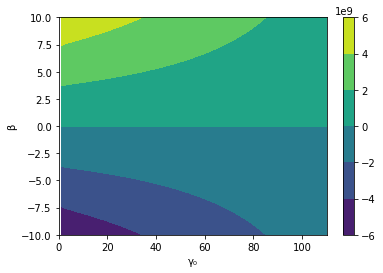}
   \caption{$p^{eff}$ for $k =+1$}
 \end{minipage}
 \end{figure}
 \begin{itemize}
 \item Clearly, $\rho^m>0$ for either $(\gamma_0>86.8,\beta\ge0)$ or $(\gamma_0<86.8,\beta\le0)$. On the other hand, if $\gamma_0>86.8$ and $\beta$ is negative, for a very small value of $|\beta|$, e.g., ($\gamma_0=87,\beta>-0.0033$), $\rho^m>0$. Similarly, non-negative $\rho^m$ is also obtained for $\gamma_0<86.8$ and a very small positive value of $\beta$, for example, ($\gamma_0=86.5,\beta<0.002$).
 \item $\rho^m-p^m>0$ is satisfied for either $(\gamma_0>107.3,\beta\ge0)$ or $(\gamma_0<107.3,\beta\le0)$. Non-negative $\rho^m-p^m$ is also obtained if ($\gamma_0>107.3,\beta<0$) or ($\gamma_0<107.3, \beta>0$), but for a very small $|\beta|$, for example, in the range ($\gamma_0=107.5,\beta>-0.0047$) and ($\gamma_0=107,\beta<0.0031$).
 \item $\rho^m+p^m>0$ is satisfied trivially for either $(\gamma_0>59.6,\beta\ge0)$ or $(\gamma_0<59.6,\beta\le0)$. Also, $\rho^m+p^m>0$ if $(\gamma_0>59.6,\beta<0)$ or $(\gamma_0<59.6,\beta>0)$ with very small $|\beta|$, e.g., ($\gamma_0=60,\beta>-0.00066$) and ($\gamma_0=59.5,\beta<0.0026$).
\item On a closer look at the expression (\ref{seck=1}), $\rho^m+3p^m>0$ requires a positive $\beta$ for any $\gamma_0>0$. For example, we see that for $\gamma_0=0.05, \beta>0.000024$ and for $\gamma_0=0.01$, $\beta>4.45\times 10^{-6}$ in account for $\rho^m+3p^m>0$. As we consider larger value of $\gamma_0$, we need comparatively smaller $\beta$ for a non-negative $\rho^m+3p^m$.
\end{itemize}
   
\paragraph{Open Universe model, $k=-1$:}
Continuing with the case $k=-1$, we first rewrite (\ref{wecp})--(\ref{secp}) and in a similar fashion determine the positive zeroes (at $86.8, 107.3$ and $59.6$, respectively) of the below polynomial coefficients of $\beta$, excluding the SEC.
\begin{align}
    \rho^m=&\frac{\beta  (80912.7 \gamma_0^4+6.19588\times 10^6 \gamma_0^3-1.14815\times 10^9 \gamma_0^2+3.88829\times 10^7 \gamma_0-242738.0)}{\gamma_0^2}+13828.2\\
    \rho^m-p^m=&\frac{\beta  (94052.4 \gamma_0^4+4.88109\times 10^6 \gamma_0^3-1.60729\times 10^9 \gamma_0^2+5.74778\times 10^7 \gamma_0-373443.0)}{\gamma_0^2}+23509.1\\
     \rho^m+p^m=&\frac{\beta  (67773.0 \gamma_0^4+7.51068\times 10^6 \gamma_0^3-6.89017\times 10^8 \gamma_0^2+2.02879\times 10^7 \gamma_0-112033.0)}{\gamma_0^2}+4147.37\\
    \rho^m+3p^m=&\frac{\beta  (41493.7 \gamma_0^4+1.01403\times 10^7 \gamma_0^3+2.29256\times 10^8 \gamma_0^2-1.69021\times 10^7 \gamma_0+149377.0)}{\gamma_0^2}-15214.4.\label{seck-1}
\end{align}
 Below are the findings:
\begin{itemize}
    \item $\rho^m>0$ is satisfied trivially for either $(\gamma_0>86.8,\beta\ge0)$ or $(\gamma_0<86.8,\beta\le0)$. It is also true if $(\gamma_0>86.8,\beta<0)$ or ($\gamma_0<86.8,\beta>0$) but $|\beta|$ is very small, e.g., ($\gamma_0=87,\beta>-0.0037$) and ($\gamma_0=86.5,\beta<0.0022$).
     \item For either $(\gamma_0>107.3,\beta\ge0)$ or $(\gamma_0<107.3,\beta\le0)$, $\rho^m-p^m>0$ is trivially satisfied. It is also satisfied if ($\gamma_0>107.3,\beta<0$) or ($\gamma_0<107.3,\beta>0$), for a small $|\beta|$, for example, ($\gamma_0=107.5,\beta>-0.0048$) and ($\gamma_0=107,\beta<0.0031$).
    \item $\rho^m+p^m>0$ is satisfied for either $(\gamma_0>59.6,\beta\ge0)$ or $(\gamma_0<59.6,\beta\le0)$. If ($\gamma_0>59.6,\beta<0$) or ($\gamma_0<59.6,\beta>0$), then $|\beta|$ has to be very small, for example, ($\gamma_0=60,\beta>-0.0007$) and ($\gamma_0=59.5,\beta<0.0022$), to satisfy this condition. 
    \item From equation (\ref{seck-1}), $\rho^m+3p^m>0$ is satisfied for all range of $\beta$ when $0<\gamma_0<0.01$, $0.01<\gamma_0<0.063$ and $\gamma_0>0.063$. For examples, ($\gamma_0=0.005, \beta>5.38755\times 10^{-6}$), ($\gamma_0=0.05, \beta<-0.00031$) and ($\gamma_0=0.07, \beta>0.0008$).
\end{itemize}
The results are illustrated in the following pictures.
\begin{figure}[H]
 \begin{minipage}[b]{0.4\textwidth}
   \includegraphics[width=\textwidth]{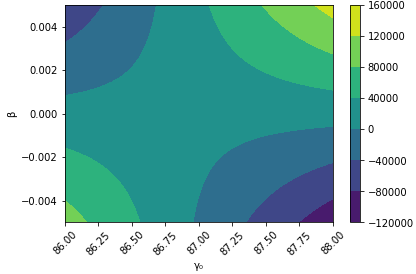}
   \caption{$\rho^m$ for $k=-1$}
 \end{minipage}
 \hfill
 \begin{minipage}[b]{0.4\textwidth}
   \includegraphics[width=\textwidth]{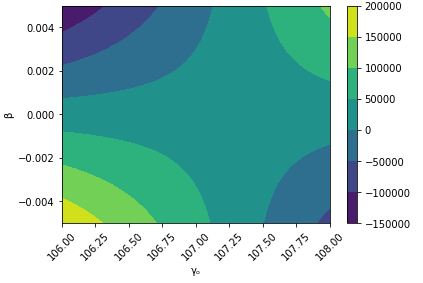}
   \caption{$\rho^m-p^m$ for $k=-1$}
 \end{minipage}
 \begin{minipage}[b]{0.4\textwidth}
   \includegraphics[width=\textwidth]{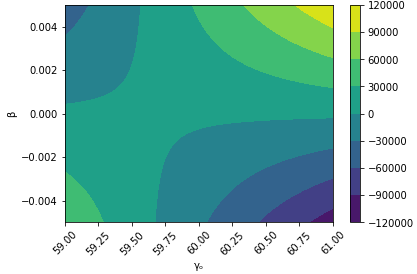}
   \caption{$\rho^m+p^m$ for $k=-1$}
 \end{minipage}
 \quad\quad\quad\quad\quad\quad\quad\quad\quad\quad\begin{minipage}[b]{0.4\textwidth}
   \includegraphics[width=\textwidth]{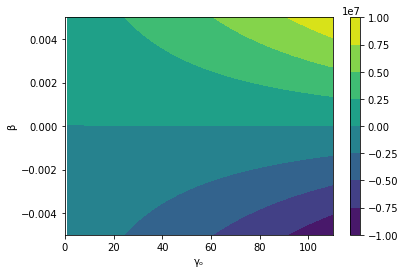}
   \caption{$\rho^m+3p^m$ for $k=-1$}
 \end{minipage}
\end{figure}
 \begin{figure}[H]
  \hfill
\begin{minipage}[b]{0.45\textwidth}
  \includegraphics[width=\textwidth]{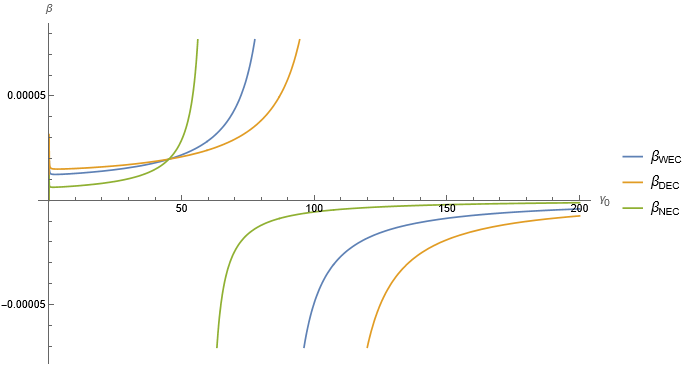}
   \caption{$\beta_{WEC},\beta_{DEC}, \beta_{NEC}$ vs $\gamma_0$ for $k =-1$}
 \end{minipage}
 \quad\quad\quad\quad\quad\quad\quad\quad\begin{minipage}[b]{0.4\textwidth}
   \includegraphics[width=\textwidth]{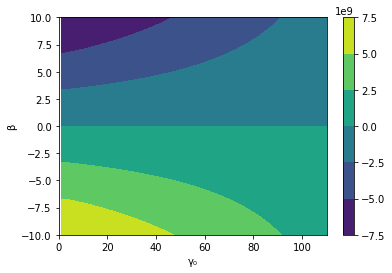}
   \caption{$p^{eff}$ for $k=-1$}
 \end{minipage}
\end{figure}
\newpage
\subsubsection{\textbf{$\gamma(t)\propto a(t)$}}
In this subsection, we consider $\gamma(t)=c_3 a(t)$ where $c_3$ is some positive constant.

\paragraph{Closed Universe model, $k=1$:} Equations (\ref{wecp})--(\ref{secp}) can be expressed as
\begin{align}
    \rho^m=&\frac{\beta  (37344.3 c_3^4+1.74705\times 10^7 c_3^3-1.14749\times 10^9 c_3^2-2.76083\times 10^7 c_3-203319.0)}{c_3^2}+13834.2\\
    \rho^m-p^m=&\frac{\beta  (-5532.49 c_3^4+2.40436\times 10^7 c_3^3-1.60660\times 10^9 c_3^2-3.08022\times 10^7 c_3-165975.0)}{c_3^2}+23517.1\\
    \rho^m+p^m=&\frac{\beta  (80221.1 c_3^4+1.08973\times 10^7 c_3^3-6.88370\times 10^8 c_3^2-2.44144\times 10^7 c_3-240663.0)}{c_3^2}+4151.37\\
    \rho^m+3p^m=&\frac{\beta  (165975.0 c_3^4-2.24905\times 10^6 c_3^3+2.29862\times 10^8 c_3^2-1.80266\times 10^7 c_3-315352.0)}{c_3^2}-15214.4.
\end{align}
We analyze the ECs in the neighbourhood of the zeroes of the polynomial ($c_3= 58.4, 67.9$ and $4278, 46.9$) and observe the following:
\begin{itemize}
    \item $\rho^m>0$ is satisfied for either $(c_3>58.4,\beta\ge0)$ or $(c_3<58.4,\beta\le0)$. It is also satisfied if ($c_3>58.4,\beta>0$) or ($c_3<58.4,\beta>0$), for a very small $|\beta|$ (e.g., $c_3=58.5,\beta>-0.0074$ and $c_3=58,\beta<0.0015$).
    \item $\rho^m-p^m>0$ is satisfied for either $(c_3>67.9,\beta\ge0)$ or $(c_3<67.9,\beta\le0)$. If ($c_3>67.9,\beta<0$) or ($c_3<67.9,\beta>0$), then $|\beta|$ has to be very small (e.g., $c_3=68,\beta>-0.01$ and $c_3=67.5,\beta<0.0025$). In addition,  $\rho^m-p^m>0$ is satisfied for either ($c_3>4278,\beta\le0$) or $(c_3<4278,\beta\ge0)$. If ($c_3>4278,\beta>0$) or ($c_3<4278,\beta<0$), then $|\beta|$ has to be very small (e.g., $c_3=4278.5,\beta<0.0020$ and $c_3=4277.5,\beta>-0.002$).     
    \item $\rho^m+p^m>0$ is satisfied for either $(c_3>46.9,\beta\ge0)$ or $(c_3<46.9,\beta\le0)$. If ($c_3>46.9,\beta<0$) or ($c_3<46.9,\beta>0$), then $|\beta|$ has to be very small for that purpose (e.g. $c_3=47,\beta>-0.0084$ and $c_3=46.5,\beta<0.000047$).
    \item $\rho^m+3p^m>0$ is satisfied for the range of $\beta$ when $0<c_3<0.093$ and $c_3>0.093$. For examples, ($c_3=0.05, \beta<-0.000059$) and ($c_3=0.1, \beta>0.00085$).
\end{itemize}
The results are shown in the following pictures.
\begin{figure}[H]
 \begin{minipage}[b]{0.4\textwidth}
   \includegraphics[width=\textwidth]{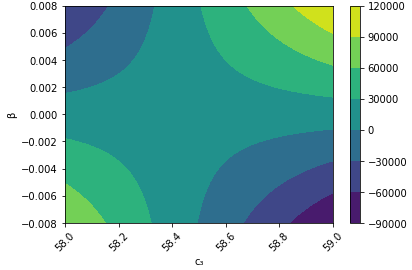}
   \caption{$\rho^m$ for $k=+1$}
 \end{minipage}
 \hfill
 \begin{minipage}[b]{0.4\textwidth}
   \includegraphics[width=\textwidth]{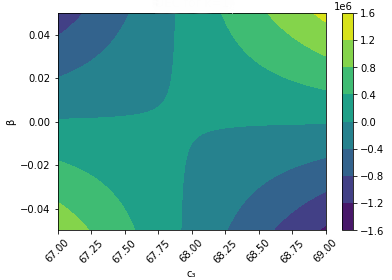}
   \caption{$\rho^m-p^m$ for $k=+1$}
 \end{minipage}
 \end{figure}
 \begin{figure}[H]
 \begin{minipage}[b]{0.4\textwidth}
   \includegraphics[width=\textwidth]{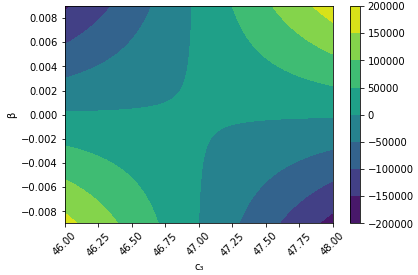}
   \caption{$\rho^m+p^m$ for $k=+1$}
 \end{minipage}
 \hfill
 \begin{minipage}[b]{0.4\textwidth}
   \includegraphics[width=\textwidth]{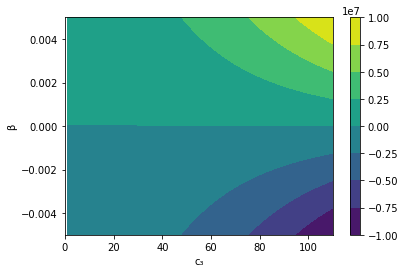}
   \caption{$\rho^m+3p^m$ for $k=+1$}
 \end{minipage}
  \hfill
 \begin{minipage}[b]{0.45\textwidth}
  \includegraphics[width=\textwidth]{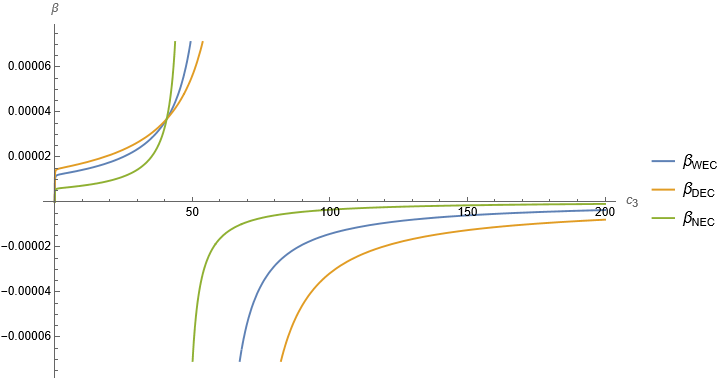}
   \caption{$\beta_{WEC},\beta_{DEC}, \beta_{NEC}$ vs $c_3$ for $k =+1$}
 \end{minipage}
 \hfill
 \begin{minipage}[b]{0.4\textwidth}
   \includegraphics[width=\textwidth]{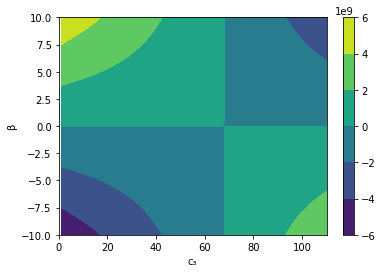}
   \caption{$p^{eff}$ for $k=+1$}
 \end{minipage}
\end{figure}
\paragraph{Open Universe model, $k=-1$:} Equations (\ref{wecp})--(\ref{secp}) can be expressed as 
\begin{align}
    \rho^m=&\frac{\beta  (37344.3 c_3^4+1.74656\times 10^7 c_3^3-1.14815\times 10^9 c_3^2+2.76132\times 10^7 c_3-203319.0)}{c_3^2}+13828.2\\
    \rho^m-p^m=&\frac{\beta  (-5532.49 c_3^4+2.40404\times 10^7 c_3^3-1.60729\times 10^9 c_3^2+3.08054\times 10^7 c_3-165975.0)}{c_3^2}+23509.1\\
    \rho^m+p^m=&\frac{\beta  (80221.1 c_3^4+1.08908\times 10^7 c_3^3-6.89011\times 10^8 c_3^2+2.44209\times 10^7 c_3-240663.0)}{c_3^2}+4147.37\\
    \rho^m+3p^m=&\frac{\beta  (165975.0 c_3^4-2.25883\times 10^6 c_3^3+2.29265\times 10^8 c_3^2+1.80364\times 10^7 c_3-315352.0)}{c_3^2}-15214.4.
\end{align}
A similar analysis yields the following information: 
\begin{itemize}
    \item $\rho^m>0$ is satisfied for either $(c_3>58.4,\beta\ge0)$ or $(c_3<58.4,\beta\le0)$. It is also satisfied for negative $\beta$ when $c_3>58.4$, but $|\beta|$ must be very small (e.g., $c_3=58.5,\beta>-0.0074$). Similarly, it is also satisfied if $c_3<58.4$ and $\beta$ is small positive, for example, ($c_3=58,\beta<0.0015$).
    \item $\rho^m-p^m>0$ is satisfied for either $(c_3>67.9,\beta\ge0)$ or $(c_3<67.9,\beta\le0)$. It is also satisfied if ($c_3>67.9,\beta<0$) or ($c_3<67.9,\beta>0$), with small $|\beta|$ (e.g., $c_3=68,\beta>-0.01$ and $c_3=67.5,\beta<0.0025$). In addition, the condition is also satisfied  for either $(c_3>4277.4,\beta\le0)$ or $(c_3<4277.4,\beta\ge0)$. If ($c_3>4277.4,\beta>0$) or ($c_3<4277.4$ and $\beta<0$), then $|\beta|$ has to be very small (e.g. $c_3=4277.5,\beta<0.0095$) and ($c_3=4277,\beta>-0.0025$).
    \item $\rho^m+p^m>0$ is satisfied for either $(c_3>46.9,\beta\ge0)$ or $(c_3<46.9,\beta\le0)$. It is also satisfied if ($c_3>46.9,\beta<0$) or ($c_3<46.9,\beta>0$), for a very small $|\beta|$ (e.g. $c_3=47,\beta>-0.007$ and $c_3=46.5,\beta<0.00048$).
    \item $\rho^m+3p^m>0$ is satisfied for the range of $\beta$ when the range of $0<c_3<0.0147$ and $c_3>0.0147$. For examples, ($c_3=0.01, \beta<-0.000013$) and ($c_3=0.16, \beta>0.000046$).
\end{itemize}
The results are depicted in the following pictures.
\begin{figure}[H]
 \begin{minipage}[b]{0.4\textwidth}
   \includegraphics[width=\textwidth]{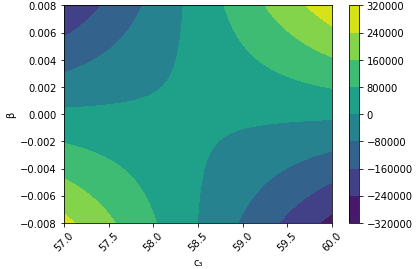}
   \caption{$\rho^m$ for $k=-1$}
 \end{minipage}
 \hfill
 \begin{minipage}[b]{0.4\textwidth}
   \includegraphics[width=\textwidth]{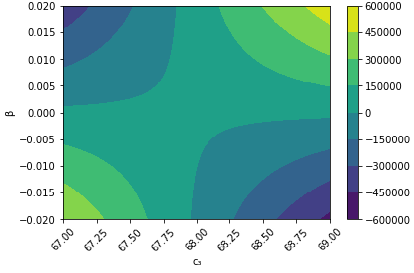}
   \caption{$\rho^m-p^m$ for $k=-1$}
 \end{minipage}
 \begin{minipage}[b]{0.4\textwidth}
   \includegraphics[width=\textwidth]{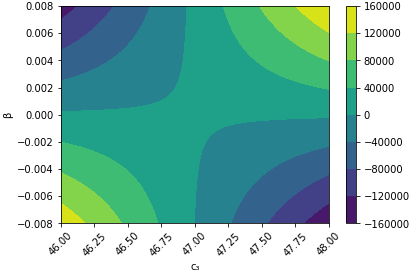}
   \caption{$\rho^m+p^m$ for $k=-1$}
 \end{minipage}
 \hfill
 \begin{minipage}[b]{0.4\textwidth}
   \includegraphics[width=\textwidth]{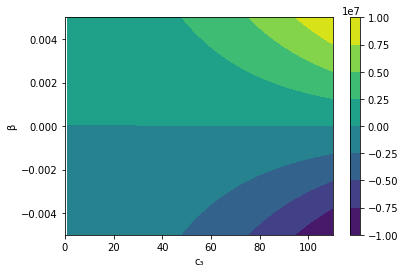}
   \caption{$\rho^m+3p^m$ for $k=-1$}
 \end{minipage}
  \end{figure}
 \begin{figure}[H]
 \hfill
  \begin{minipage}[b]{0.45\textwidth}
  \includegraphics[width=\textwidth]{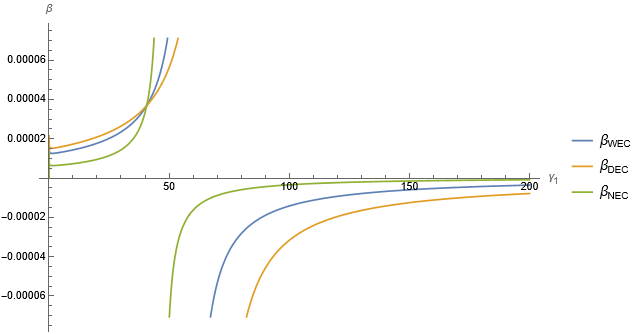}
   \caption{$\beta_{WEC},\beta_{DEC}, \beta_{NEC}$ vs $c_3$ for $k =-1$}
 \end{minipage}
 \hfill
   \begin{minipage}[b]{0.4\textwidth}
   \includegraphics[width=\textwidth]{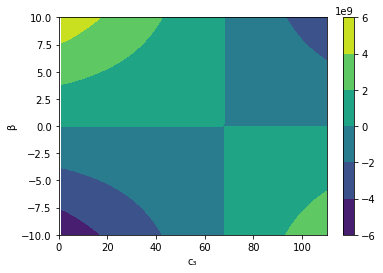}
   \caption{$p^{eff}$ for $k=-1$}
 \end{minipage}
\end{figure}


\newpage
\subsection{$f(Q)= Q+\beta \sqrt{-Q}$}\label{model2}
In this subsection, we consider this particularly fascinating $f(Q)$ model which was studied a number of time in the literature. It emerged initially in a spatially flat FLRW under a coincident gauge to mimic a $\Lambda$CDM evolution in GR \cite{cosmology_Q}. The model was tested against redshift space distortion data, likelihood \cite{redshift} and Bayesian analysis \cite{siren}, matter power spectrum and lensing effect on the Cosmic Microwave Background radiation (CMB) angular power spectrum \cite{signature}, the dynamics of the linear matter perturbations and gravitational potentials \cite{lcdm1}, among others. A reasonable range for the coefficient $\beta$ was highly sought after and debatable till date. Although the model is not contributing except a boundary term in the spatially flat FLRW background in coincident gauge, once we broaden our spectra to include the spatial curvature, it is worthwhile to investigate the model and look at the range of $\beta$ from the energy condition standpoint. 

For this model, the energy condition expressions in terms of the present values $H_0$ and $q_0$ can be written in a similar way as

\begin{align}
    \rho^m=&\frac{\beta\sqrt{-Q_0}}{4}+3\left(1+\frac{\beta}{2\sqrt{-Q_0}}\right)\left(\frac{k}{a_0^2}+H_0^2\right)+\frac{3\beta\dot{Q_0}}{8(\sqrt{-Q_0})^3}\left(\frac{k}{\gamma}+\frac{\gamma}{a_0^2}\right)\label{wecq2}\\
    \rho^m-p^m=&\frac{\beta\sqrt{-Q_0}}{2}+2\left(1+\frac{\beta}{2\sqrt{-Q_0}}\right)\left(\frac{2k}{a_0^2}+(2-q_0)H_0^2\right)-\frac{\beta\dot{Q_0}}{2(\sqrt{-Q_0})^3}\left(H_0-\frac{\gamma}{a_0^2}\right)\label{decq2}\\
    \rho^m+p^m=&2\left(1+\frac{\beta}{2\sqrt{-Q_0}}\right)\left((1+q_0)H_0^2+\frac{k}{a_0^2}\right)+\frac{\beta\dot{Q_0}}{4(\sqrt{-Q_0})^3}\left(2H_0+\frac{3k}{\gamma}+\frac{\gamma}{a_0^2}\right)\label{necq2}\\
    \rho^m + 3p^m=&-\frac{\beta}{2}\sqrt{-Q_0}+6\left(1+\frac{\beta}{2\sqrt{-Q_0}}\right)q H_0^2+\frac{3\beta\dot{Q_0}}{2(\sqrt{-Q_0})^3}\left(H_0+\frac{k}{\gamma}\right)\label{secq2}
\end{align}
where $Q_0$ and $\dot{Q_0}$ are given previously in equation (\ref{Q}).  

\subsubsection{$\gamma(t)=\gamma_0$}
\paragraph{Closed Universe model, $k=1$:} Equations (\ref{wecq2})--(\ref{secq2}) can be expressed as follows

\begin{align}
    \rho^m=&-\frac{0.121357 \beta  \left(6.61535 \gamma _0^4+2194.27 \gamma _0^3+427.742 \gamma _0^2-13766.3 \gamma _0-284.460\right)}{\sqrt{\gamma _0} \left(3+135.771 \gamma _0-\gamma _0^2\right)^{\frac 32}}+13834.2\label{wecqc1}\\
    \rho^m-p^m=&-\frac{0.121357 \beta  \left(-10.7806 \gamma _0^4+3523.45 \gamma _0^3+675.171 \gamma _0^2-25739.3 \gamma _0-529.228\right)}{\sqrt{\gamma _0} \left(3+135.771 \gamma _0-\gamma _0^2\right)^{\frac 32}}+23517.1\label{decqc1}\\
    \rho^m+p^m=&-\frac{0.121357 \beta  \left(24.0113 \gamma _0^4+865.092 \gamma _0^3+180.313 \gamma _0^2-1793.27 \gamma _0-39.6921\right)}{\sqrt{\gamma _0} \left(3+135.771 \gamma _0-\gamma _0^2\right)^{\frac 32}}+4151.37\label{necqc1}\\
    \rho^m+3p^m=&-\frac{0.121357 \beta  \left(58.8031 \gamma _0^4-1793.27 \gamma _0^3-314.546 \gamma _0^2+22152.7 \gamma _0+449.844\right)}{\sqrt{\gamma _0} \left(3+135.771 \gamma _0-\gamma _0^2\right)^{\frac 32}}-15214.4.\label{secqc1}
\end{align}

At a quick glance, it is clear that we require ($\gamma_0< 135.79$) for real values of pressure and energy density. Now for the respective ECs to be satisfied, we can proceed to find the suitable ranges of the model parameter $\beta$:
\begin{itemize}
    \item Suppose we express the equation (\ref{wecqc1}) as $\rho^m=13834.2-0.121357\beta B$, where 
    $$B=\frac{\left(6.61535 \gamma _0^4+2194.27 \gamma _0^3+427.742 \gamma _0^2-13766.3 \gamma _0-284.460\right)}{\sqrt{\gamma _0} \left(3+135.771 \gamma _0-\gamma _0^2\right)^{\frac 32}}.$$
    The denominator of $B$ is positive for $0<\gamma_0<135.79$, numerator is positive for $\gamma_0>2.411$ and negative for $0<\gamma_0\leq 2.411$. Therefore, $\beta<\frac{13834.2}{0.121358B}$ for $2.411<\gamma_0<135.79$ and $\beta>\frac{13834.2}{0.121358B}$ for $0<\gamma_0\leq 2.411$ are required for a positive $\rho^m$. For example, ($\gamma_0=0.5, \beta>-7051.74$) and ($\gamma_0=135, \beta<0.19$).
    \item Proceeding in a similar manner, from (\ref{decqc1}) we can find a suitable dynamic range of $\beta$ for any given value of $\gamma_0<2.62$ and $2.62<\gamma_0<135.79$ to get non-negative $\rho^m-p^m>0$.
    \item In a similar way, $\rho^m+p^m>0$ is satisfied for $\beta$ when the range of $\gamma_0<1.32$ and $1.32<\gamma<135.79$. For examples, ($\gamma_0=1,\quad \beta>-72448.3$) and ($\gamma_0=135,\quad \beta<-0.043$).
    \item Similarly, for $\rho^m+3p^m>0$ we can find suitable range of $\beta$ for any $\gamma_0<3.65$, $3.65<\gamma_0<30.26$ and $30.26<\gamma_0<135.79$. For examples, ($\gamma_0=10, \beta>17489.1$) and ($\gamma_0=135, \beta<-0.10$).
\end{itemize}
The results are shown in the following pictures.
\begin{figure}[H]
 \begin{minipage}[b]{0.4\textwidth}
   \includegraphics[width=\textwidth]{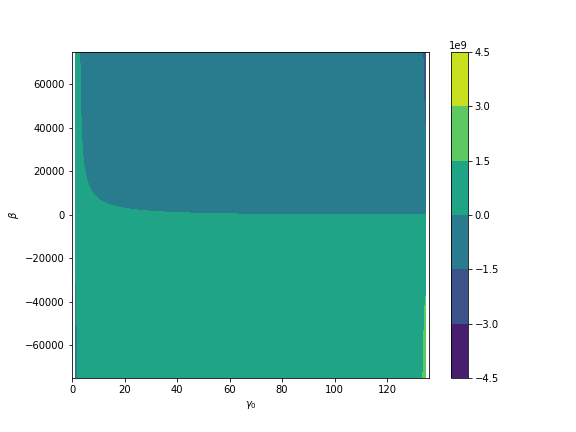}
   \caption{$\rho^m$ for k =+1}
 \end{minipage}
 \hfill
 \begin{minipage}[b]{0.4\textwidth}
   \includegraphics[width=\textwidth]{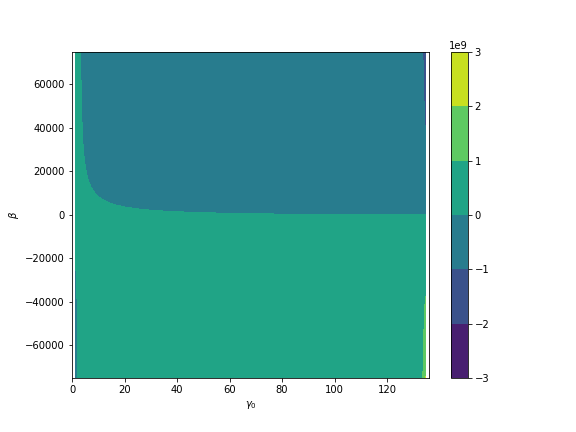}
   \caption{$\rho^m-p^m$ for k =+1}
 \end{minipage}
 \hfill
 \begin{minipage}[b]{0.4\textwidth}
   \includegraphics[width=\textwidth]{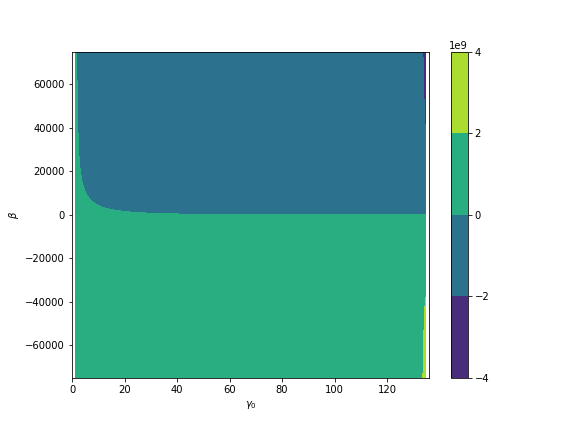}
   \caption{$\rho^m+p^m$ for k = +1}
 \end{minipage}
 \hfill
 \begin{minipage}[b]{0.4\textwidth}
   \includegraphics[width=\textwidth]{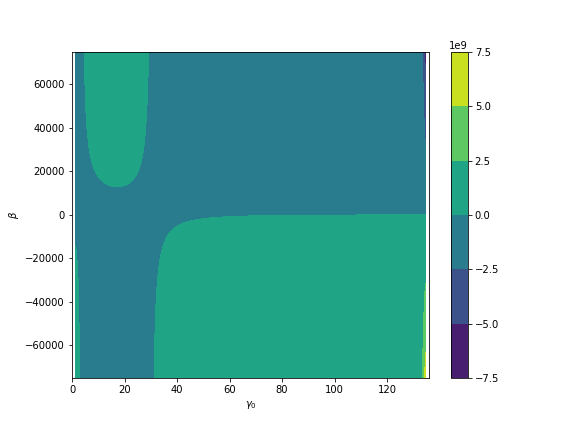}
   \caption{$\rho^m+3p^m$ for k = +1}
 \end{minipage}
\end{figure}
\begin{figure}[H]
 \begin{minipage}[b]{0.45\textwidth}
  \includegraphics[width=\textwidth]{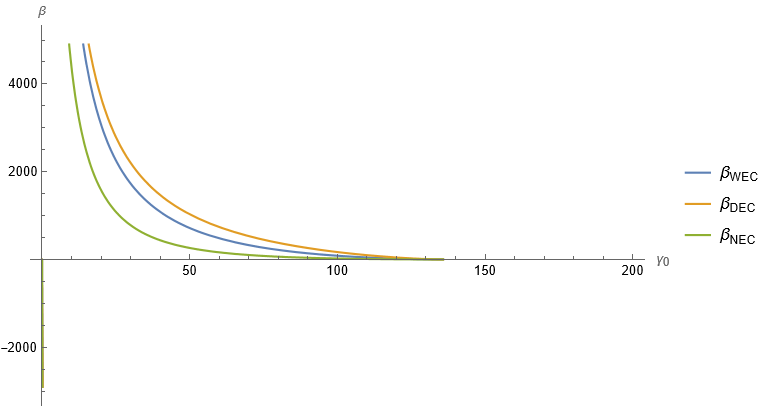}
   \caption{$\beta_{WEC},\beta_{DEC}, \beta_{NEC}$ vs $\gamma_0$ for $k =+1$}
 \end{minipage}
 \hfill
  \begin{minipage}[b]{0.4\textwidth}
   \includegraphics[width=\textwidth]{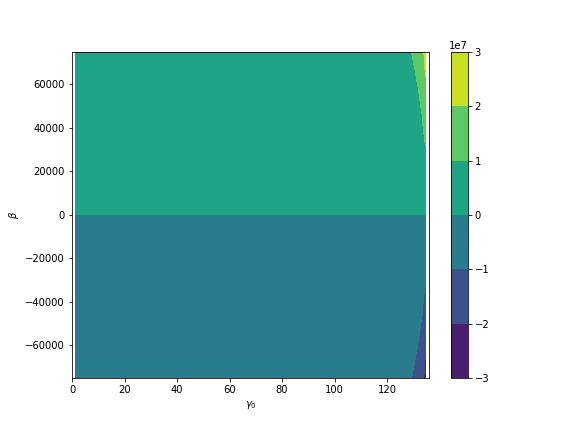}
   \caption{$p^{eff}$ for k = +1}
 \end{minipage}
\end{figure}

\paragraph{Open Universe model, $k=-1$:} Equations (\ref{wecq2})--(\ref{secq2}) can be expressed as follows

\begin{align}
    \rho^m=&-\frac{0.121357 \beta  \left(6.61535 \gamma _0^4+2197.73 \gamma _0^3-427.844 \gamma _0^2+13783.6 \gamma _0-284.460\right)}{\sqrt{\gamma _0} \left(-\gamma _0^2+135.829 \gamma _0-3\right)^{\frac{3}{2}}}+13828.2\label{wecq221}\\
    \rho^m-p^m=&-\frac{0.121357 \beta  \left(-10.7806 \gamma _0^4+3530.38 \gamma _0^3-675.341 \gamma _0^2+25767.0 \gamma _0-529.228\right)}{\sqrt{\gamma _0} \left(-\gamma _0^2+135.829 \gamma _0-3\right)^{\frac{3}{2}}}+23509.1\label{decq221}\\
    \rho^m+p^m=&-\frac{0.121357 \beta  \left(24.0113 \gamma _0^4+865.092 \gamma _0^3-180.347 \gamma _0^2+1800.19 \gamma _0-39.6921\right)}{\sqrt{\gamma _0} \left(-\gamma _0^2+135.829 \gamma _0-3\right)^{\frac{3}{2}}}+4147.37\label{necq221}\\
    \rho^m+3p=&-\frac{0.121357 \beta  \left(58.8031 \gamma _0^4-1800.19 \gamma _0^3+314.648 \gamma _0^2-22166.6 \gamma _0+449.844\right)}{\sqrt{\gamma _0} \left(-\gamma _0^2+135.829 \gamma _0-3\right)^{\frac{3}{2}}}-15214.4.\label{secq221}
\end{align}
Naturally, $0.022<\gamma_0<135.8$ is required to make the sum of the terms under $\sqrt{.} \,$ non-negative. Following the similar arguments as for the case $k=+1$, we find the suitable dynamic ranges of $\beta$ for each EC to be satisfied:
\begin{itemize}
    \item For $\rho^m>0$, $\beta$ has to be $\beta<\frac{13828.2}{0.121357A}$ where $$A=\frac{\left(6.61535 \gamma _0^4+2197.73 \gamma _0^3-427.844 \gamma _0^2+13783.6 \gamma _0-284.460\right)}{\sqrt{\gamma _0} \left(-\gamma _0^2+135.829 \gamma _0-3\right)^{\frac{3}{2}}}.$$ For example, given $\gamma_0=135$, one needs $\beta<0.19$ for the $\rho^m>0$ to be satisfied.\\
    \item Similarly, for example, say $\gamma_0=135$, then the range of $\beta$ has to be $\beta<0.50$ for the $\rho^m-p^m<0$ to be satisfied.\\
    \item In the same manner, say $\gamma_0=135$, then the range of $\beta$ has to be $\beta<0.044$ for the $\rho^m+p^m>0$ to be satisfied.\\
    \item Likewise, we can find suitable $\beta$ for the range $0.022<\gamma_0<30.83$ and $30.84<\gamma_0<135.8$ for the $\rho^m+3p^m>0$ to be satisfied. For examples, ($\gamma_0=0.5, \beta>4277.7$) and ($\gamma_0=135, \beta<-0.109$).  
\end{itemize}
The results of these analyses are depicted in the following figures.
\begin{figure}[H]
 \begin{minipage}[b]{0.4\textwidth}
   \includegraphics[width=\textwidth]{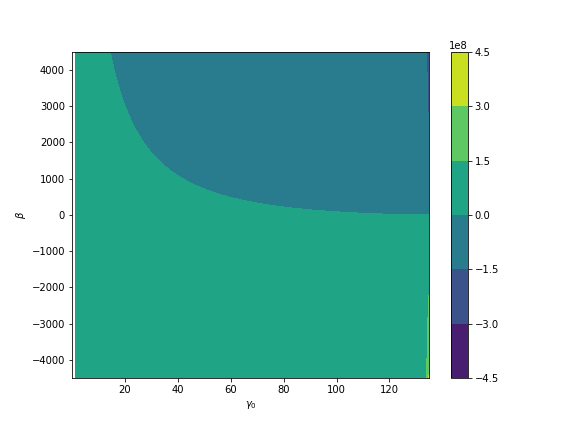}
   \caption{$\rho^m$ for k =-1}
 \end{minipage}
 \hfill
 \begin{minipage}[b]{0.4\textwidth}
   \includegraphics[width=\textwidth]{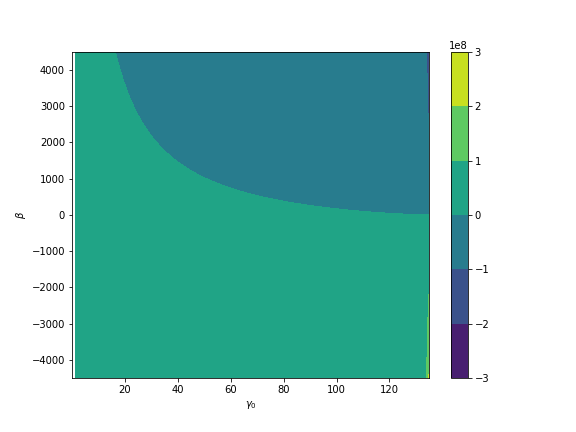}
   \caption{$\rho^m-p^m$ for k = -1}
 \end{minipage}
 \hfill
 \begin{minipage}[b]{0.4\textwidth}
   \includegraphics[width=\textwidth]{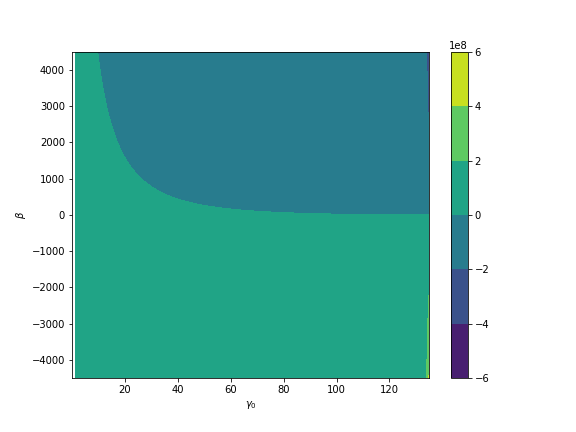}
   \caption{$\rho^m+p^m$ for k = -1}
 \end{minipage}
 \hfill
 \begin{minipage}[b]{0.4\textwidth}
   \includegraphics[width=\textwidth]{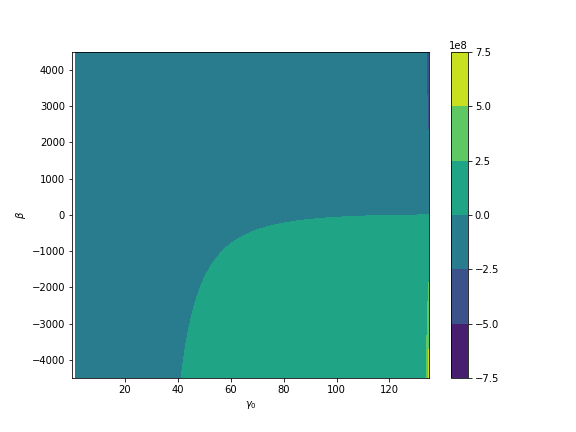}
   \caption{$\rho^m+3p^m$ for k = -1}
 \end{minipage}
  \hfill
 \begin{minipage}[b]{0.45\textwidth}
  \includegraphics[width=\textwidth]{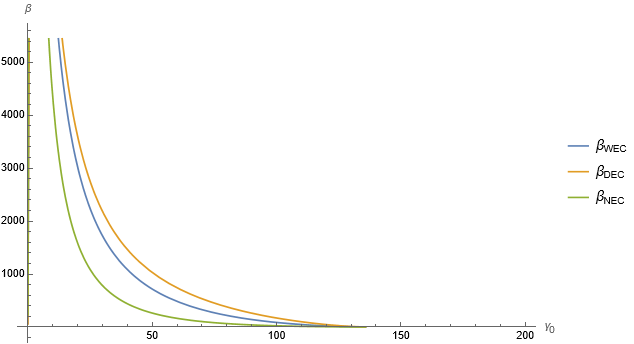}
   \caption{$\beta_{WEC},\beta_{DEC}, \beta_{NEC}$ vs $\gamma_0$ for $k =-1$}
 \end{minipage}
 \hfill
 \begin{minipage}[b]{0.4\textwidth}
   \includegraphics[width=\textwidth]{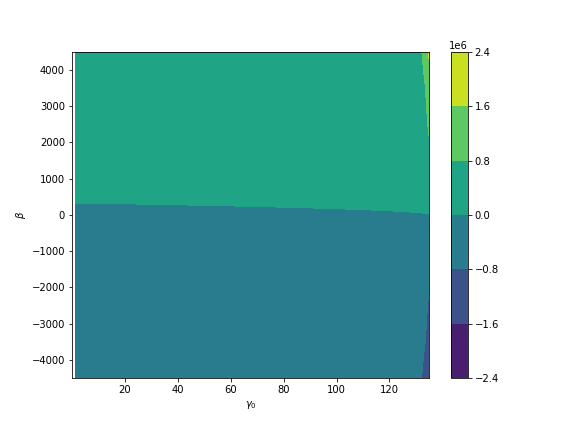}
   \caption{$p^{eff}$ for k = -1}
 \end{minipage}
\end{figure}

\subsubsection{$\gamma(t)\propto a(t)$}
In this subsection, we consider $\gamma(t)=\gamma_1 a(t)$ where $\gamma_1$ is some positive constant.
\paragraph{Closed Universe model, $k=1$:} Equations (\ref{wecq2})--(\ref{secq2}) can be expressed as
\begin{align}
    \rho^m=&-\frac{0.0858124 \beta  \left(-37.4870 \gamma _1^4+3092.20 \gamma _1^3+235.187 \gamma _1^2-4888.07 \gamma _1-80.1193\right)}{\sqrt{\gamma _1} \left(-\gamma _1^2+67.8853 \gamma _1+1\right)^{\frac{3}{2}}}+13834.2\label{wecq22}\\
    \rho^m-p^m=&-\frac{0.0858124 \beta  \left(-89.1847 \gamma _1^4+6051.60 \gamma _1^3+381.198 \gamma _1^2-7248.84 \gamma _1-117.606\right)}{\sqrt{\gamma _1} \left(-\gamma _1^2+67.8853 \gamma _1+1\right)^{\frac{3}{2}}}+23517.1\label{decq22}\\
    \rho^m+p^m=&-\frac{0.0858124 \beta  \left(14.2108 \gamma _1^4+132.802 \gamma _1^3+89.1762 \gamma _1^2-2527.29 \gamma _1-42.6323\right)}{\sqrt{\gamma _1} \left(-\gamma _1^2+67.8853 \gamma _1+1\right)^{\frac{3}{2}}}+4151.37\label{necq22}\\
    \rho^m+3p^m=&-\frac{0.0858124 \beta  \left(117.606 \gamma _1^4-5786.0 \gamma _1^3-202.845 \gamma _1^2+2194.27 \gamma _1+32.3417\right)}{\sqrt{\gamma _1} \left(-\gamma _1^2+67.8853 \gamma _1+1\right)^{\frac{3}{2}}}-15214.4.\label{secq22}
\end{align}

\begin{figure}[H]
 \begin{minipage}[b]{0.4\textwidth}
   \includegraphics[width=\textwidth]{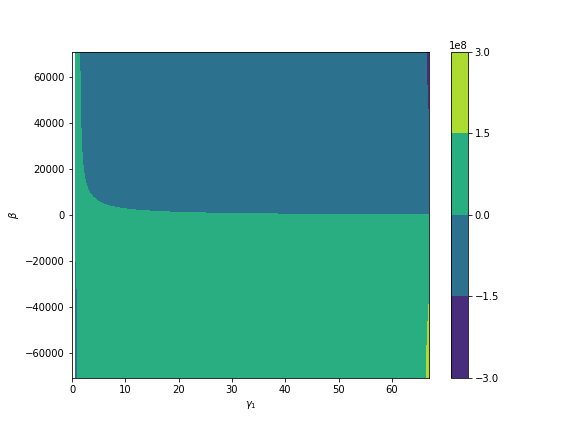}
   \caption{$\rho^m$ for k =+1}\label{fig37}
 \end{minipage}
 \hfill
 \begin{minipage}[b]{0.4\textwidth}
   \includegraphics[width=\textwidth]{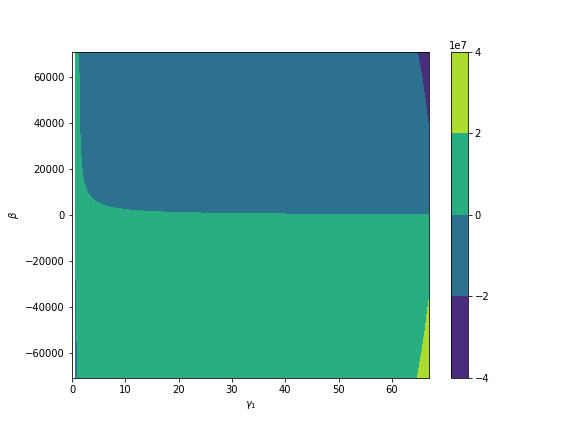}
   \caption{$\rho^m-p^m$ for k = +1}\label{fig38}
 \end{minipage}
  \begin{minipage}[b]{0.4\textwidth}
   \includegraphics[width=\textwidth]{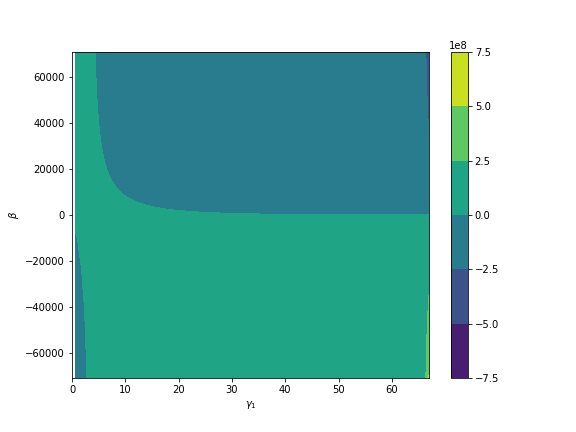}
   \caption{$\rho^m+p^m$ for k = +1}\label{fig39}
 \end{minipage}
 \hfill
 \begin{minipage}[b]{0.4\textwidth}
   \includegraphics[width=\textwidth]{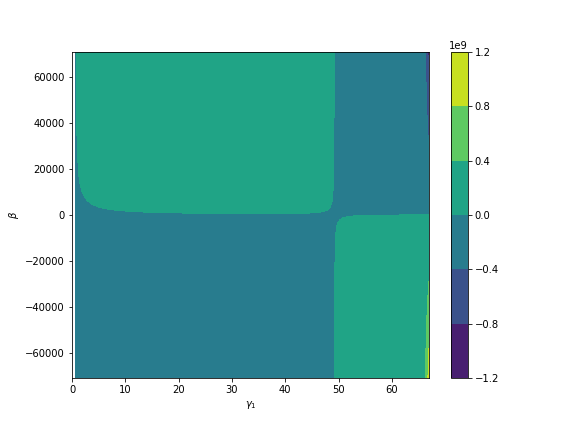}
   \caption{$\rho^m+3p^m$ for k = +1}\label{fig40}
 \end{minipage}
 \hfill
 \begin{minipage}[b]{0.45\textwidth}
  \includegraphics[width=\textwidth]{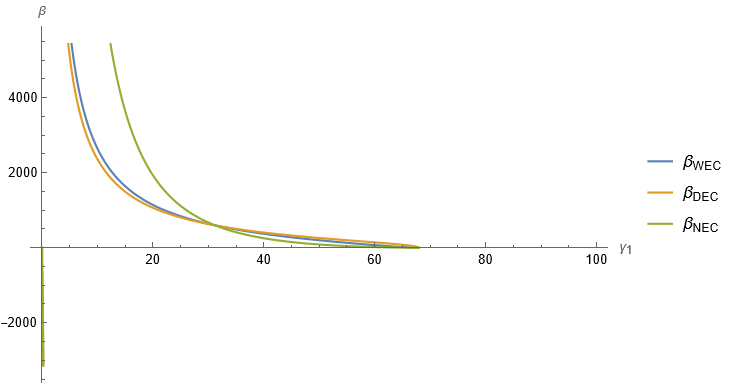}
   \caption{$\beta_{WEC},\beta_{DEC}, \beta_{NEC}$ vs $\gamma_1$ for $k =+1$}
 \end{minipage}
 \hfill
 \begin{minipage}[b]{0.4\textwidth}
   \includegraphics[width=\textwidth]{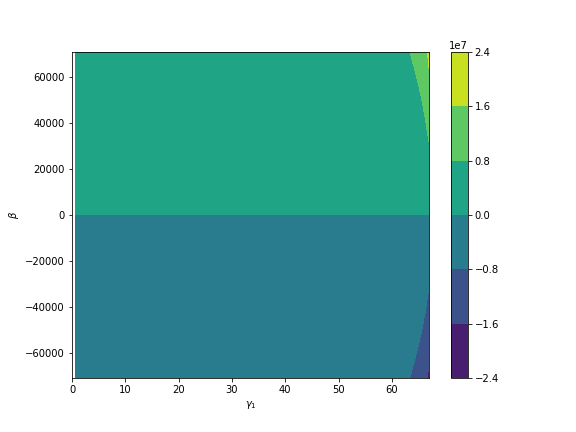}
   \caption{$p^{eff}$ for k = +1}
 \end{minipage}
\end{figure}

First we observe that $\gamma_1<67.9$ is required to avoid imaginary components in our discussion. The respective ranges of $\beta$ for each ECs are discussed below:
\begin{itemize}
    \item For simplicity, equation (\ref{wecq22}) can be expressed as $\rho^m=13834.2-0.0858124 \beta A$ where 
    $$A=\frac{\left(-37.4870 \gamma _1^4+3092.20 \gamma _1^3+235.187 \gamma _1^2-4888.07 \gamma _1-80.1193\right)}{\sqrt{\gamma _1} \left(-\gamma _1^2+67.8853 \gamma _1+1\right)^{\frac{3}{2}}}.$$ The denominator of $A$ is positive for $0<\gamma_1<67.9$, the numerator is negative for $0<\gamma_1<1.23$ and positive for $\gamma_1>1.23$. Hence, to satisfy $\rho^m>0$, $\beta$ has to be $\beta>\frac{13834.2}{0.085814 A}$ for the range of $0<\gamma_1<1.23$ and for $1.23<\gamma_1<67.9$, $\beta>\frac{13834.2}{0.085814 A}$. For examples, ($\gamma_1=1, \beta>-53728.3$) and ($\gamma_1=67, \beta<3.52$). The result is depicted in Figure (\ref{fig37}).\\
    \item In a similar manner, from equation (\ref{decq22}), $\rho^m-p^m>0$ is satisfied for the range of $\beta$ when the range of $\gamma_1<1.07$ and $1.07<\gamma_1<67.9$. For examples, ($\gamma_1=1, \beta>-149862$) and ($\gamma_1=67, \beta<43.5$). The result is depicted in Figure (\ref{fig38}).\\
    \item On the other hand, from equation (\ref{necq22}), $\rho^m+p^m>0$ is satisfied for the range of $\beta$ when $0<\gamma_1<3.49$ and $3.49<\gamma_1<67.9$. For examples, ($\gamma_1=2, \beta>-30332$) and ($\gamma_1=67, \beta<0.56$). The result is depicted in Figure (\ref{fig39}).\\
    \item Furthermore, from equation (\ref{secq22}), $\rho^m+3p^m>0$ is satisfied for the range of $\beta$ when $0<\gamma_1<0.609$, $0.609<\gamma_1<49.2$ and $\gamma_1>49.2$. For examples, ($\gamma_1=0.5, \beta<-70598.8$), ($\gamma_1=49, \beta>11216.9$) and ($\gamma_1=67, \beta<-1.08$). The result is depicted in Figure (\ref{fig40}).
\end{itemize}
The results are illustrated in the following pictures.

\paragraph{Open Universe model, $k=-1$:} Equations (\ref{wecq2})--(\ref{secq2}) can be expressed as

\begin{align}
    \rho^m=&-\frac{0.0858124 \beta  \left(-37.4870 \gamma _1^4+3096.53 \gamma _1^3-235.238 \gamma _1^2+4894.13 \gamma _1-80.1193\right)}{\sqrt{\gamma _1} \left(-\gamma _1^2+67.9147 \gamma _1-1\right)^{\frac{3}{2}}}+13828.2\label{wecq222}\\
    \rho^m-p^m=&-\frac{0.0858124 \beta  \left(-89.1847 \gamma _1^4+6059.69 \gamma _1^3-381.283 \gamma _1^2+7258.08 \gamma _1-117.606\right)}{\sqrt{\gamma _1} \left(-\gamma _1^2+67.9147 \gamma _1-1\right)^{\frac{3}{2}}}+23509.1\label{decq222}\\
    \rho^m+p^m=&-\frac{0.0858124 \beta  \left(14.2108 \gamma _1^4+133.380 \gamma _1^3-89.1932 \gamma _1^2+2530.17 \gamma _1-42.6323\right)}{\sqrt{\gamma _1} \left(-\gamma _1^2+67.9147 \gamma _1-1\right)^{\frac{3}{2}}}+4147.37\label{necq222}\\
    \rho^m+3p=&-\frac{0.0858124 \beta  \left(117.606 \gamma _1^4-5792.93 \gamma _1^3+202.896 \gamma _1^2-2197.73 \gamma _1+32.3417\right)}{\sqrt{\gamma _1} \left(-\gamma _1^2+67.9147 \gamma _1-1\right)^{\frac{3}{2}}}-15214.4\,.\label{secq222}
\end{align}
Clearly, $0.015<\gamma_1<67.9$ is required in this scenario for the coefficient of $\beta$, we offer the following analysis:
\begin{itemize}
    \item In equation (\ref{wecq222}), the range of $\beta$ has to be $\beta>\frac{13828.2}{0.0858124 A}$ for $0.0150<\gamma_1<0.0164$ and $\beta<\frac{13828.2}{0.0858124 A}$ for $0.0164<\gamma_1<67.9$ where $$A=\frac{\left(-37.4870 \gamma _1^4+3096.53 \gamma _1^3-235.238 \gamma _1^2+4894.13 \gamma _1-80.1193\right)}{\sqrt{\gamma _1} \left(-\gamma _1^2+67.9147 \gamma _1-1\right)^{\frac{3}{2}}}$$ for the $\rho^m>0$ to be satisfied. For examples, ($\gamma_1=0.016, \beta>-278.1$) and ($\gamma_1=67, \beta<3.52$).\\
    \item Similarly, $\rho^m-p^m>0$ is satisfied for range of $\beta$ when $0.0150<\gamma_1<0.0162$ and $0.0163<\gamma_1<67.9$. For examples, ($\gamma_1=0.016, \beta>-567.7$) and ($\gamma_1=67, \beta<43.4$).\\
    \item In a similar manner, $\rho^m+p^m>0$ is satisfied for the range of $\beta$ when $0.0150<\gamma_1<0.0168$ and $0.0169<\gamma_1<67.9$. For examples, ($\gamma_1=0.016,\beta>-71.46$) and ($\gamma_1=67, \beta<0.567$).\\
    \item Likewise, $\rho^m+3p^m>0$ is satisfied for the range of $\beta$ when $0.015<\gamma_1<49.2$ and $49.3<\gamma_1<67.9$. For examples, ($\gamma_1=0.016, \beta>203.79$) and ($\gamma_1=67, \beta<-1.081$).
\end{itemize}
The results are depicted in the following figures.
\begin{figure}[H]
 \begin{minipage}[b]{0.4\textwidth}
   \includegraphics[width=\textwidth]{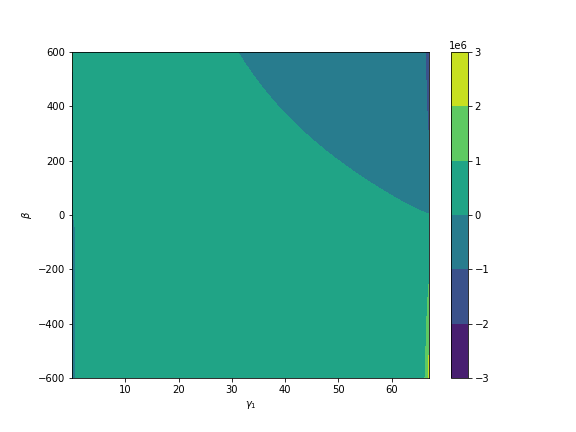}
   \caption{$\rho^m$ for k =-1}
 \end{minipage}
 \hfill
 \begin{minipage}[b]{0.4\textwidth}
   \includegraphics[width=\textwidth]{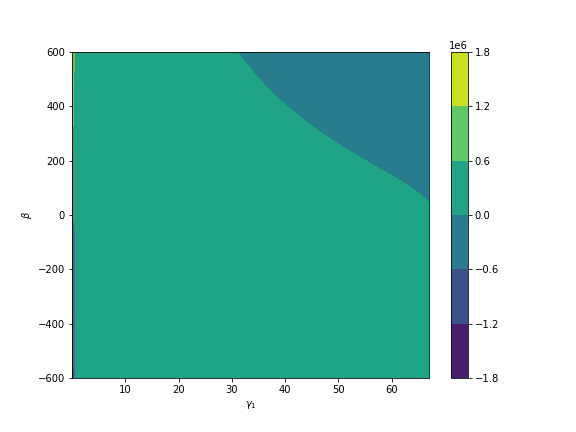}
   \caption{$\rho^m-p^m$ for k = -1}
 \end{minipage}
 \end{figure}
 \begin{figure}[H]
 \begin{minipage}[b]{0.4\textwidth}
   \includegraphics[width=\textwidth]{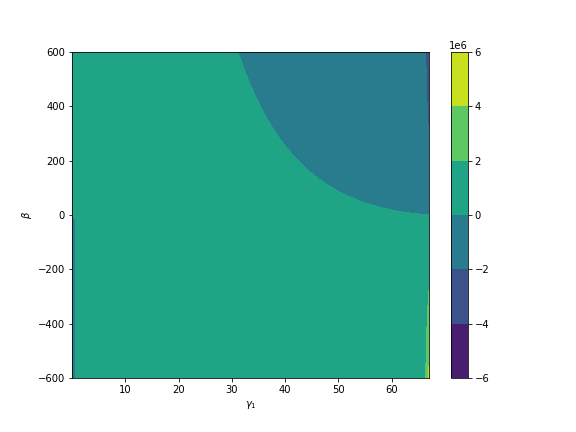}
   \caption{$\rho^m+p^m$ for k = -1}
 \end{minipage}
 \hfill
 \begin{minipage}[b]{0.4\textwidth}
   \includegraphics[width=\textwidth]{f=Q+beta (-Q)^1by2/model1_case_2_k-1NEC.png}
   \caption{$\rho^m+3p^m$ for k = -1}
 \end{minipage}
\end{figure}
\begin{figure}[H]
  \hfill
 \begin{minipage}[b]{0.45\textwidth}
  \includegraphics[width=\textwidth]{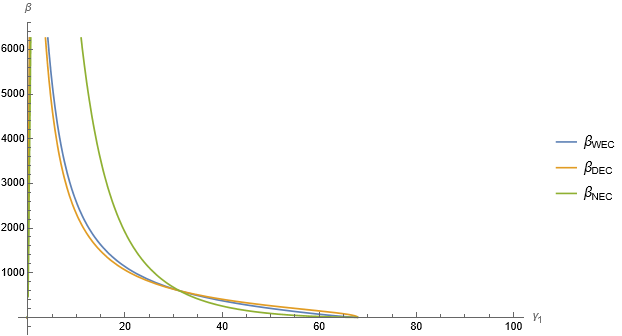}
   \caption{$\beta_{WEC},\beta_{DEC}, \beta_{NEC}$ vs $\gamma_1$ for $k =-1$}
 \end{minipage}
 \hfill
 \begin{minipage}[b]{0.4\textwidth}
   \includegraphics[width=\textwidth]{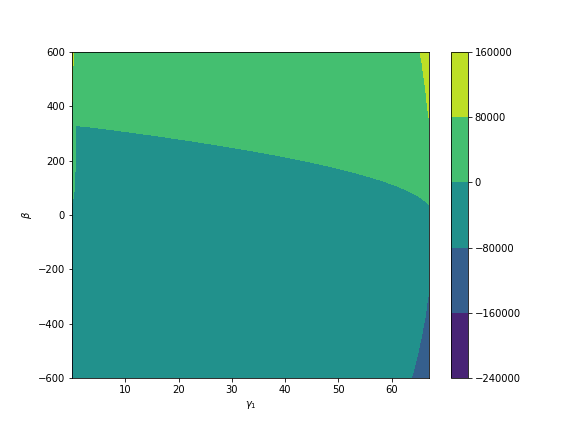}
   \caption{$p^{eff}$ for k = -1}
 \end{minipage}
\end{figure}



\section{Concluding remarks}\label{sec6}
A thorough analysis of the energy conditions in $f(Q)$ theory in open and closed type FLRW Universe has been offered here. The formulation of the $f(Q)$ theory in such background spacetime depends on an unknown and so far unrestricted time-varying parameter $\gamma(t)$. In our study, we have considered some reasonable ansatz, $\gamma(t)=\gamma_0$, a constant, and $\gamma(t)\propto a(t)$, to be examined from the perspective of energy conditions.
Considering ordinary barotropic fluid as the matter source, we have derived the Friedmann-like equations of pressure and energy density for the ordinary matter and their effective counterparts. Two of the most popular $f(Q)$ models, namely, $f(Q)=Q+\beta Q^2$ and $f(Q)=Q+\beta \sqrt{-Q}$ have been considered and the range of ($\beta,\gamma$) for valid ECs are computed. Observational values of some cosmological parameters have been used for this purpose, yielding an effective equation of state $\omega^{eff}=-0.7$. For each scenario, we have extensively analyzed the range of the free parameters, from the ECs of the ordinary matter as well as put forth the corresponding ranges of negative effective pressure $p^{eff}$, which is an essential component of modified gravity accounting for the late-time acceleration without depending on the physical existence of DE.


\end{document}